# Spin polarized transport current in n-type co-doped ZnO thin films measured by Andreev spectroscopy

Karen A Yates
The Blackett Laboratory, Physics Department, Imperial College London, SW7 2AZ, UK
Anthony J Behan, James R Neal, David.S. Score, Harry.J Blythe and Gillian A Gehring
Department of Physics and Astronomy, University of Sheffield, S3 7RH, UK
Steve. M. Heald
Argonne National Laboratory, Argonne IL 60439, USA
Will.R. Branford and Lesley F Cohen
The Blackett Laboratory, Physics Department, Imperial College London, SW7 2AZ, UK

We use point contact Andreev reflection measurements to determine the spin polarization of the transport current in pulse laser deposited thin films of ZnO with 1% Al and with and without 2%Mn. Only films with Mn are ferromagnetic and show spin polarization of the transport current of up to 55 ± 0.5% at 4.2 K, in sharp contrast to measurements of the nonmagnetic films without Mn where the polarization is consistent with zero. Our results imply strongly that ferromagnetism in these Al doped ZnO films requires the presence of Mn.

74.45+c, 72.25.Dc, 75.50.Pp

The realization of a semiconducting system with spin polarized carriers and above room temperature Curie temperature ($T_c$) has the potential to lead to a new generation of spintronic devices with revolutionary electrical and magnetic properties [1]. One material of great interest is the doped ZnO system for which ferromagnetism has been observed at room temperature [2,3]. The origin of ferromagnetism is still open to wide fundamental speculation because in some cases, ferromagnetism is observed without there being a moment on the dopant transition metal ion itself [4] and also because similar ferromagnetic behaviour is observed when ZnO is doped with a non-magnetic element such as carbon [5]. These materials are characterized by unusually high Curie temperatures coupled with low magnetic moments and may be described in terms of a narrow split band of localised [6] or mobile carriers [7] (for $CaB_6$ in the latter case). These systems are not considered to be dilute magnetic semiconductors in the usual sense but have been proposed to exhibit a new and unusual form of magnetism [8].

An understanding of these materials was further extended by recent experiments showing that, when ZnO is heavily doped to beyond the Mott limit for metallic conduction ($n_c \geq 1 \times 10^{20} cm^{-3}$), the saturation magnetization, $M_{sat}$, shows a dependence on the carrier concentration and magnetic dopant density [9,10] characteristic of the model proposed by Chattopadhyay et al, [11] for the GaMnAs system. Nonetheless, this cannot be the complete picture as the dependence of $M_{sat}$ on $n_c$ still shows a large amount of scatter [9]. These results have led some authors to claim that structural or strain effects may be an essential component for the appearance of FM [12] and there is further evidence suggesting that grain boundaries play an important role [13,14].

Independently from this current work, ferromagnetism in the Mn doped ZnO system has been confirmed by spectroscopic studies including EPR [15,16], X-ray circular dichroism, which probes the magnetic state of the Mn ions [17], and optical circular dichroism which probes the magnetism of the valance and the conduction bands [9, 18]. In this paper, we show that the Andreev data provides a definitive observation of spin polarized carriers in the bulk of the Mn doped ZnO co-doped with Al in marked contrast to films doped only with Al (ZAO). The measurements support a temperature independence of the polarization (P) in the range of temperatures studied, consistent with the temperature independence of the magnetic moment.

The films used in this study were grown by pulsed laser deposition (PLD) as described earlier, [9,10]. The films were deposited onto c-plane sapphire and were oriented. The ZnO films were doped with nominally 1% Al, (films B and C) or 1% Al and 2% Mn (film A). The films were grown to be between 120-240nm thick. Film A was characterized at beamline 20-BM at the Advanced Photon Source. X-ray fluorescence data determined the actual Mn concentration to be 3.8 ± 0.3 %. In previous studies, where the TM dopant concentration was found to be higher than nominal, this was attributed to Zn deficiency during the PLD process [19,20]. As seen in Figure 1, both the EXAFS and the XANES data are consistent with all of the Mn residing on tetrahedral Zn sites. The edge position in Fig. 1(a) is consistent with $Mn^{2+}$. The strong pre-edge peak near 6540eV is also typical of tetrahedral bonding. Furthermore, the distinct double peaked pre-edge that would result from $Mn^{3+}$ in a tetrahedral site [21] was not seen, giving additional confidence that the Mn was predominantly in the 2+ state, as required for magnetism [22]. While a small amount (5%) of secondary phase cannot be definitely excluded, the near neighbour distance and coordination determined by EXAFS is consistent with all of the Mn residing in tetrahedral substitutional sites. As shown in Fig. 1(b), the Fourier transform of the Mn EXAFS is very similar to a similar transform for the Zn EXAFS from pure ZnO. Also shown in Fig 1(b) is a fit to the Mn EXAFS for the first three shells of atoms assuming Mn in a substitutional Zn site. The reduction in the peak amplitudes is found to be due to increased disorder likely caused by the lattice distortion needed to accommodate the larger $Mn^{2+}$ ion. The near neighbor Mn-O distance is about 2.08 Å; this is 0.1 Å larger then the Zn-O distance and similar to that expected for $Mn^{2+}$ in a tetrahedral site.

In order to be measured using PCAR, the films were heavily co-doped with Al to beyond the Mott limit and therefore showed metal-like conduction with resistance ratios slightly greater than one [9]. Resistivity as a function of temperature is shown in figure 2a for a ZAO film (film C). Magnetization loops at 5K (figure 2b) indicate that, when co-doped with Mn, the ZAO film is ferromagnetic with a coercive field of 160 Oe and a saturation magnetization, $M_{sat}$, of 0.4 $\mu_B$/Mn (inset to figure 2). The pure ZAO films showed no ferromagnetism at any temperature.

Point contact spectra were obtained using a mechanically sharpened Nb tip as described in [23]. PCAR spectra were fitted according to the Blonder-Tinkham-Klapwijk (BTK) model [24] modified to include spin polarization by Mazin et al., [25]. The important parameters of the fit are the superconducting energy gap, $\Delta$, the spin polarization, P, the interface parameter, Z, and the broadening (including thermal broadening) parameter, $\omega$. We have previously shown that such a four parameter fit is potentially degenerate and a unique value for the transport spin transport polarization is only obtained when there is a minimum in the "$\chi^2(P)$" fitting procedure as outlined in ref [23]. Briefly, a trial value of P, $P_{trial}$, is used to calculate the least squares fit to the data using the remaining three parameters. By plotting the $\chi^2$ as a function of P, a minimum in $\chi^2(P)$ is obtained if there is a unique solution for the transport spin polarization. If on the other hand, thermal broadening or inelastic scattering adds smearing to the spectrum such that the effects of P and those of Z on the conductance spectra compensate for each other [26], a divergence is obtained in the $\chi^2(P)$ which indicates an upper limit of the polarization for that contact, $P_{upper}$ [23]. This is shown in the inset to figure 3 where a spectrum that can be fitted with a transport spin polarization of 46% is compared with a spectrum that shows a divergent $\chi^2(P)$ and therefore can only be fitted with an upper limit P of ~ 30%.

Spectra were fitted assuming a ballistic transport regime. From the range of resistivity values and mean free paths for the films, the contact size was estimated (using the Sharvin formula) to be between $6.7 \leq d \leq 9.5$nm, compared with a mean free path ($\lambda$) of between 0.5 and 2.9nm. Although $d$ is of the same order of magnitude as $\lambda$, the fact that the contact is normally composed of many nano-contacts means that our assumption is likely to be robust.

The results of the fitting of a large number of different contacts is shown in figure 3. Both types of film (ie.ZAO films with and without co-doping ) show a distribution of P with interface parameter, Z, as has been observed previously in many different systems [27]. The key difference here is that high polarization values are *only* obtained for film A, ie. the film that is doped with both Mn and Al. For this film, contacts showing P values of up to 55 ± 0.5% were obtained indicating a significant, intrinsic, transport spin polarization. In contrast the pure ZAO films showed a low distribution of P values with an upper value of ~ $P_{upper} \leq 15\%$. It is important to note however, that the Andreev method used at these temperatures cannot distinguish between $P_{upper}$ of 15-20% and P = 0%, as has been shown previously using Cu as the reference sample [23,28]. Hence, the fact that Al doped films show $P_{upper}$ ~ 15% is indeed consistent with an unpolarized transport current.

The effect of contact pressure on P, shown in figure 4, supports our statement that ZnO thin films doped only with Al are unpolarized. In general, as the sharpened tip comes into contact with the film, the conductance of the contact ($1/R_N$) increases. This can be caused by two effects, either the contact size can increase, or the tip can puncture a more resistive surface layer resulting in a decrease in overall contact resistance ($1/R_N$ increases). For the case of the ZnO films doped only with Al, although there is a wide change in $1/R_N$ with tip pressure, P remains well below $P_{upper}$ ~ 15%. In contrast, for ZnO co-doped with Al and Mn, high values of polarization (and low values of Z) are observed at large values of $1/R_N$. The data suggests that the increase in P as $1/R_N$ increases is associated with the tip breaking through a surface layer in the film as the tip pressure is increased. The general trend cannot be explained by simply changing the contact size as the pressure on the tip is changed. However, there is one anomalous data point (for film A in figure 4) and this most probably relates to precisely this effect. An increase in tip area would increase the number of nanochannels, thereby increasing $1/R_N$ without changing P. The high values of P in figure 4 are then the intrinsic bulk spin polarization of the ferromagnetic ZnO co-doped with Al and Mn.

We have also studied the effect of temperature as is shown in figure 5. As the temperature is increased, so the spectra broaden and the features due to Z and P are smeared out. The $\chi^2(P)$ fitting procedure however, adequately captures the increased smearing and shows that while the smearing increases linearly with temperature, P remains constant and $\Delta$ decreases in a manner consistent with BCS theory (Fig. 6).

These results show quite clearly that, within the temperature range measured here, there is a spin polarized transport current in the Mn doped ZAO thin film consistent with previous reports of (Ga,Mn)As [29] and (In,Mn)Sb [30]. In contrast, the polarization of the transport current in ZAO films without Mn co-doping and which show no ferromagnetism at any temperature is consistent with that of an unpolarized metal, such as Cu. The origin of the ferromagnetism in TM doped ZnO, especially if the ZnO is highly n-type, is a matter of considerable debate, for a review see [31]. Moreover, issues of phase segregation and impurity contribution to the magnetic moment are always of considerable concern [32], however the results presented here bring significant clarity. Unlike many of the studies that show phase separation and particularly nanoparticulate or spinel contributions to the magnetism [33,34], this study was performed using Mn as the TM dopant. Unlike Co, nano-particles of Mn are not expected to be ferromagnetic and, of the various phases of $Mn_xO_y$, only $Mn_3O_4$ and $Mn_2O_3$ show ferromagnetic properties with $T_c$s of 43K and 83K respectively [18]. However, it has also been suggested that spinel phases such as $ZnMn_2O_4$ [35] and $Mn_{2-x}Zn_xO_{3-y}$ [36] or even uncompensated spins in antiferromagnetic particles could be enough to cause the apparent observation of ferromagnetism in such thin films [34]. Although the temperature dependence of the magnetization, the absence of a low energy band edge characteristic of Mn oxide phases in the MCD [9] as well as the evidence from the EXAFS and XANES data, would suggest that such secondary phases are unlikely to be the cause of the ferromagnetism observed here, the PCAR is a low temperature probe. Based on the PCAR alone therefore, we cannot make statements concerning the ferromagnetism above 10K. Nonetheless, we note that a spin polarization can only be observed using PCAR when there are *mobile* spin polarized carriers, eliminating uncompensated AFM spins as the origin of the spin polarization observed. Moreover, the fact that the majority of the spectra measured across the entire film surface show high spin polarization strongly suggests that this spin polarization is *distributed* throughout the film. The detection of a mobile, distributed, spin polarized transport current indicates that, even were unidentified nanoparticles, or isolated $Mn_xO_y$ phases, the cause of the ferromagnetism observed in the films, these nanoparticles have to polarize the transport electrons in the thin film. In summary therefore, the observation of the polarization in the ferromagnetic film supports the contention that ferromagnetism is a bulk effect.

Moreover, it is interesting to note that the 4.2K $M_{sat}$ value of 0.4$\mu_B$/Mn (or equivalently 0.014$\mu_B$/Zn) can, in fact, be accounted for by using the bulk value of P = 55% and the measured carrier density ($1.32 \times 10^{21}$ cm$^{-3}$). This observation is consistent with a model whereby a spin-split conduction band would adequately account for the low temperature magnetic data [6,7]. We cannot extrapolate P to temperatures above the $T_c$ of Nb (9.2K). Nonetheless, we do note that the saturation magnetization is roughly constant with temperature up to room temperature (figure 2), even though the spin polarization of the carriers may be expected to be reduced by this temperature [37].

In conclusion, we have shown that thin films of highly n-type ZnO co-doped with Mn and Al and which show ferromagnetic behaviour have a low temperature transport spin polarization of at least 55 ± 0.5%. In contrast, thin films of highly n-type ZnO doped with Al only and which are not ferromagnetic at any temperature, show spin transport polarization values consistent with that of Cu, ie. unpolarized. It is unphysical to attribute the results to uncompensated, localized, spins on the surfaces of antiferromagnetic nano-inclusions, implying that the ferromagnetism is only supported by the presence of Mn in these ZAO films. Interestingly, Panguluri et al, have recently demonstrated that in $In_2O_3$ the presence of the TM dopant was not a necessary component for *carrier*-mediated ferromagnetism in that system [38]. Clearly, much work remains in order to establish the role of the TM ion in carrier mediated ferromagnetism and spin polarized transport and that this may indeed be system dependent. The present data also show that any secondary magnetic phase present would have to polarize the transport current of the film to adequately explain our observations. This result has significant implications for the technological development of devices based on spintronics materials.

**Acknowledgements** This work is supported by EPSRC for work done both at Imperial College London and Sheffield. Use of the Advanced Photon Source is supported by the US Department of Energy, Office of Science, Office of Basic Energy Sciences, under contract DE-AC02-06CH11357.

**Figure captions**

Figure 1: (a) The near edge spectrum of the Mn edge in Mn doped ZnO compared to Mn standards with different valences. (b) The $k^2$-weighted Fourier transform of the Mn EXAFS (k=2-12) compared to a similar transform of the Zn EXAFS from ZnO. Also shown is a fit to the first three shells of the Mn EXAFS assuming the Mn is substituting for Zn.

Figure 2: (a) Resistivity vs temperature for film C. (b) magnetisation loops for film A at 5K (O), 20K (X), 290K (■). Inset shows an expanded view of the 5K data indicating a value for the coercive field of 160 Oe (50 Oe at 290K).

Figure 3: Polarization as a function of Z for a number of different contacts onto film A (○) and films B,C (▲, ▼). Also shown, spectra for which there is only a maximum limit for P, $P_{upper}$, for film A (X) and film C (★). Lines are guides for the eye. Inset shows the method of obtaining P and $P_{upper}$ based on the $\chi^2(P)$ method.

Figure 4: The polarization as a function of contact conductance $1/R_N$ for film A (○) and films B, C (■,▲). The arrow indicates the direction of increasing pressure while the dashed line is a guide for the eye only.

Figure 5: Spectra from film A as a function of temperature (symbols) with fits (solid lines). Spectra are shown for temperatures between 4.2K and 7K.

Figure 6: Temperature dependence of the parameters extracted from fitting the data in figure 5. (a) Δ, the line indicates expected behaviour from the BCS model, (b) P, (c) ω.

**References**

[1]  S.J. Pearton, C.R. Abernathy, D.P. Norton, AF Hebard, YD Park, LA Boatner, JD Budai, Mat Sci and Eng R: Reports, **40**, 127 (2003)
[2] S.A. Chambers, Surface Science Reports, **61**, 345 (2006)
[3] F. Pan, C. Song, X.J. Liu, YC Yang, F Zeng, Mat Sci and Engineering R : Reports, **62**, 1 (2008)
[4] T Tietze, M. Gacic, G. Schutz, G Jakob, S Bruck, E. Goering, New J Phys, 10, 055009 (2008)
[5] H. Pan, J.B. Yi, L. Shen, RQ Wu, JH Yang, JY Lin, YP Feng, J Ding, LH Van, JH Yin, Phys Rev Lett, **99**, 127201 (2007)
[6] J.M.D. Coey, M. Venkatesan, C.B. Fitzgerald, Nature Materials, **4**, 173, (2005)
[7] D.M. Edwards and M. I. Katsnelson, J. Phys: Condens Matter, **18**, 7209 (2006)
[8] M. Venkatesan, R.D. Gunning, P. Stamenov, JMD. Coey, J. Appl. Phys, **103**, 07D135 (2008)
[9] A.J. Behan, A. Mokhtari, H.J. Blythe, D Score, X-H Xu, JR Neal, AM Fox, GA Gehring, Phys. Rev. Lett, **100**, 047206 (2008)
[10] X.H. Xu, H.J. Blythe, M. Ziese, AJ Behan, JR Neal, A Mokhtari, RM Ibrahim, AM Fox, GA Gehring, New J Phys, **8**, 135 (2006)
[11] A. Chattopadhyay, S. Das Sarma, A. J. Millis, Phys Rev Lett, **87**, 227202 (2001)
[12] N.H. Hong, J. Sahai, N. Poirot and V. Brize, Phys Rev B, 73, 132404 (2006)
[13] A. Quesada, M.A. Garcia, J. De la Venta, E Fernandez Pinel, JM Merino, A. Hernando, The European Physical Journal B, Condensed Matter and Complex Systems, **59**, 457 (2007)
[14] B.B. Straumal, A.A. Mazilkin, S.G. Protasova, AA Myatiev, PB Straumal, G Schutz, PA van Aken, E Goering, B Baretzky,  Phys Rev B, 79, 205206 (2009)
[15] V.S. Dang, Y.Y. Song, N.Q. Hoa, SC Yu, YG Yoo, J Korean Phys Soc, **52**, 1398 (2008)
[16] P. Sharma, A. Gupta, K.V. Rao, FJ Owens, R Sharma, Nature Materials, **2**, 673, (2003)
[17] P. Thakur, K. H. Chae, J.-Y. Kim, M Subramanian, R Jayavel, K Asokan, Appl. Phys. Lett., **91**, 162503 (2007)
[18] J.R. Neal, A.J. Behan, R. M. Ibrahim, HJ Blythe, M Ziese, AM Fox, GA Gehring, Phys Rev. Lett, **96**, 197208 (2006)
[19] YB Zhang, Q. Liu, T Sritharan, CL Gan, S Li, Appl Phys Lett, 89, 042510, 2006
[20] M. Venkatesan, CB Fitzgerald, JG Lunney, JMD Coey, Phys Rev Lett, 93, 177206, (2004)
[21] A. Titov, X. Biquard, D. Halley, S Kuroda, E. Bellet-Amalric, H Mariette, J Cibert, AE Merad, MB Kanom, E Kulatov, Yu A Uspenskii, Physical Review B **72**, 115209 (2005)
[22] F. Bondino, K. B. Garg, E. Magnano, E Carleschi, M Heinonen, RK Singhal, SK Gaur, F Parmigiani, Journal of Physics: Condensed Matter **20**, 275205 (2008)
[23] Y. Bugoslavsky, Y. Miyoshi, S. K. Clowes, WR Branford, M Lake, I Brown, AD Caplin, LF Cohen, Phys Rev B, **71**, 104523 (2005)
[24] G. E. Blonder, M. Tinkham, and T. M. Klapwijk, Phys Rev B **25**, 4515 (1982).
[25] I. I. Mazin, A. A. Golubov, and B. Nadgorny, J. Appl. Phys. **89**, 7576 (2001)
[26] N Auth, G. Jakob, T. Block and C. Felser, Phys. Rev. B, **68**, 024403 (2003)
[27] G. T. Woods, J. Soulen, R. J., I. Mazin, B. Nadgorny, MS Osofsky, J Sanders, H Srikanth, WF Egelhoff, R Datla, Phys Rev B **70**, 054416 (2004).

[28] F. Magnus, K.A. Yates, B. Morris, Y Miyoshi, Y Bugoslavsky, LF Cohen, G Burnell, MG Blamire, PW Josephs-Franks, Appl Phys Lett, **89**, 262505 (2006)
[29] R. P. Panguluri, K. C. Ku, T. Wojtowicz, X Liu, TK Furdyna, YB Lyanda-Geller, N Samarth, B Nadgorny, Phys Rev B **72**, 054510 (2005)
[30] R. P. Panguluri, B. Nadgorny, T. Wojtowicz, X Liu, JK Furdyna, Appl Phys Lett **91**, 252502 (2007)
[31] T. Dietl, J Phys. Condens, Matter, **19**, 165204 (2007)
[32] S. Zhou, K. Potzger, J. Von Borany, R Grotzschel, W Skorupa, M Helm, J Fassbender, Phys Rev. B, **77**, 035209 (2008)
[33] M Venkatesan, P. Stamenov, L.S. Dorneles, RD Gunning, B Bernoux, JMD Coey, Appl. Phys. Lett, **90**, 242508 (2007)
[34] D. Iusan, R. Knut, B. Sanyal, O Karis, O Eriksson, VA Coleman, G Westin, J Magnus Wikberg, P Svedlindh, Phys Rev B **78**, 085319 (2008)
[35] J.H. Li, D.Z. Shen, J.Y. Zhang, DX Zhao, BS Li, YM Lu, YC Liu, XW Fan, J Magn Magn Mater, 302, 118 (2006)
[36] D. C. Kundaliya, S.B. Ogale, S.E. Lofland, S Dhar, CH Metting, SR Shinde, Z Ma, B Varughese, KV Ramanujachary, L Salamanca-Riba, T Venkatesan, Nat. Mat., 3, 709 (2004)
[37] Q. Xu, L. Hartmann, S. Zhou, A Mcklich, M Helm, G Biehne, H Hochmuth, M Lorenz, M Grundmann, H Schmidt, Physical Review Letters **101**, 076601 (2008)
[38] R.P. Panguluri, P. Kharel, C. Sudakar, R Naik, R Suryanarayanan, VM Naik, AG Petukhov, B Nadgorny, G Lawes, Phys Rev B **79**, 165208 (2009)

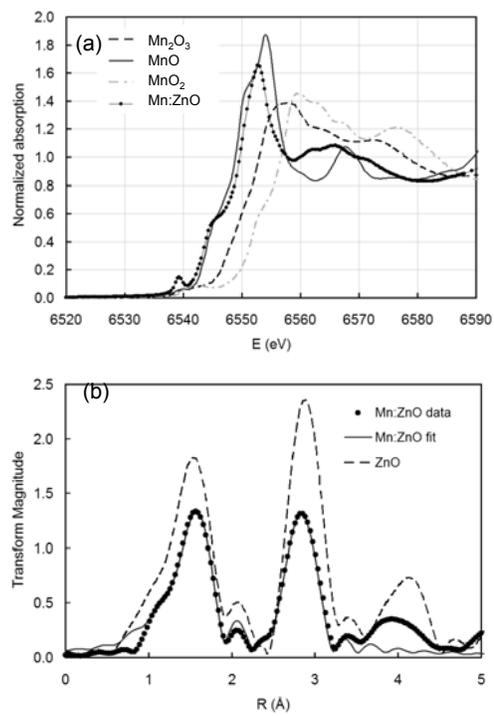

Figure 1

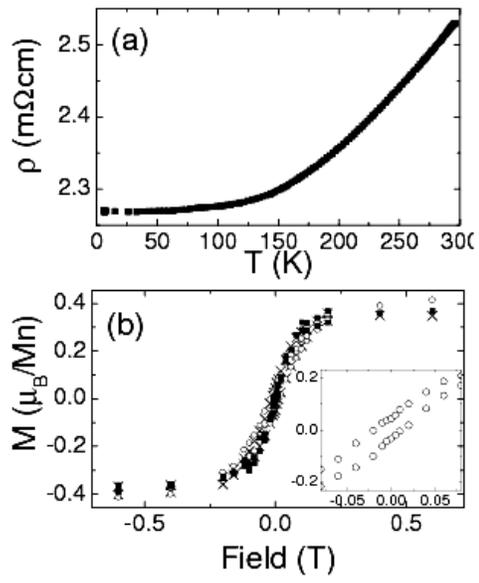

Figure 2

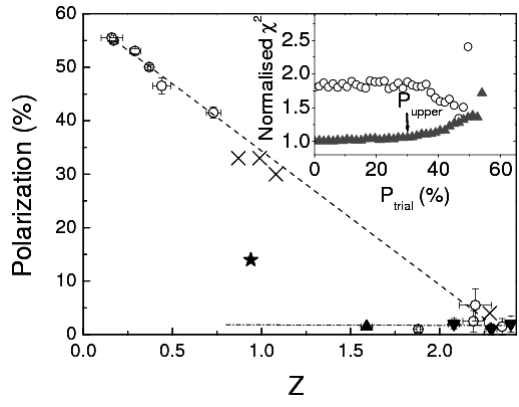

Figure 3

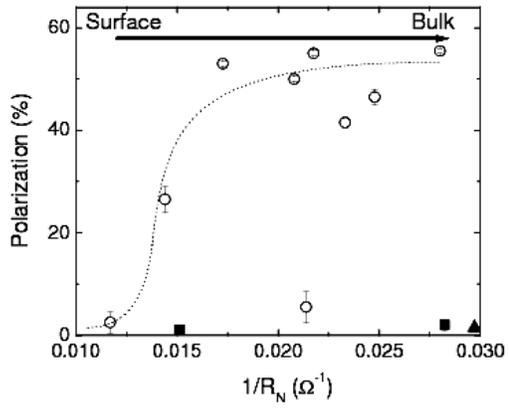

Figure 4

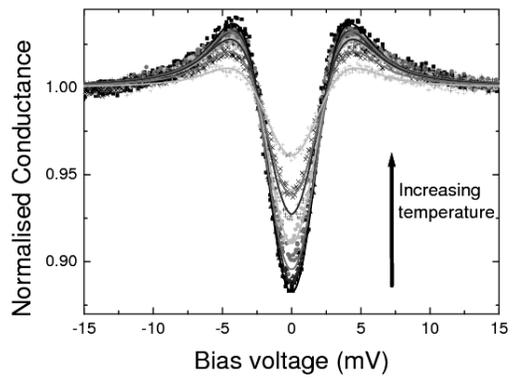

Figure 5

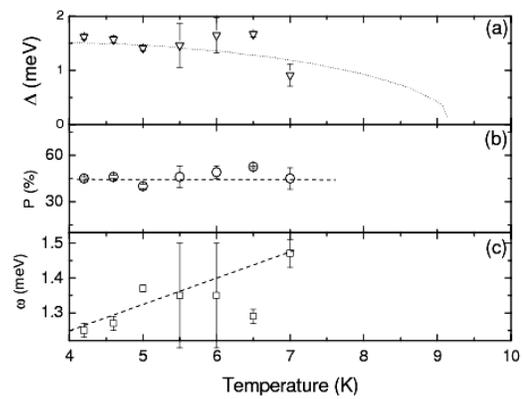

Figure 6